\documentclass[amsmath,amssymb,prl,superscriptaddress,twocolumn]{revtex4-2}

\usepackage[compact]{titlesec}
\usepackage[utf8]{inputenc}
\usepackage{graphicx}
\usepackage{dcolumn}
\usepackage{bm}
\usepackage{natbib}
\usepackage{float}
\usepackage{amsmath}
\usepackage{xcolor}

\makeatletter
\def\maketitle{
\@author@finish
\title@column\titleblock@produce
\suppressfloats[t]}
\makeatother

\newcommand{\beginsupplement}{
    \onecolumngrid
    \setcounter{page}{1}
    \setcounter{table}{0}
    \renewcommand{\thetable}{S\arabic{table}}
    \setcounter{figure}{0}
    \renewcommand{\thefigure}{S\arabic{figure}}
    \setcounter{equation}{0}
    \renewcommand{\theequation}{S\arabic{equation}}
    \setcounter{secnumdepth}{2}
}

\begin{document}

\title{Planar Josephson Junctions Templated by Nanowire Shadowing}

\author{P. Zhang}
\affiliation{Department of Physics and Astronomy, University of Pittsburgh, Pittsburgh, PA, 15260, USA}

\author{A. Zarassi}
\affiliation{Department of Physics and Astronomy, University of Pittsburgh, Pittsburgh, PA, 15260, USA}

\author{M. Pendharkar}
\affiliation{Electrical and Computer Engineering, University of California Santa Barbara, Santa Barbara, CA 93106, USA}

\author{J.S. Lee}
\affiliation{California NanoSystems Institute, University of California Santa Barbara, Santa Barbara, CA 93106, USA}

\author{L. Jarjat}
\affiliation{Département de Physique, Ecole Normale Supérieure, 75005  Paris,  France}

\author{V. Van de Sande}
\affiliation{Eindhoven University of Technology, 5600 MB, Eindhoven, The Netherlands}

\author{B. Zhang}
\affiliation{Department of Physics and Astronomy, University of Pittsburgh, Pittsburgh, PA, 15260, USA}

\author{S. Mudi}
\affiliation{Department of Physics and Astronomy, University of Pittsburgh, Pittsburgh, PA, 15260, USA}

\author{H. Wu}
\affiliation{Department of Physics and Astronomy, University of Pittsburgh, Pittsburgh, PA, 15260, USA}

\author{S. Tan}
\affiliation{Department of Electrical and Computer Engineering, University of Pittsburgh, Pittsburgh, PA 15260, USA}

\author{C.P. Dempsey}
\affiliation{Electrical and Computer Engineering, University of California Santa Barbara, Santa Barbara, CA 93106, USA}

\author{A.P. McFadden}
\affiliation{Electrical and Computer Engineering, University of California Santa Barbara, Santa Barbara, CA 93106, USA}

\author{S.D. Harrington}
\affiliation{Materials Department, University of California Santa Barbara, Santa Barbara, CA 93106, USA}

\author{B. Shojaei}
\affiliation{California NanoSystems Institute, University of California Santa Barbara, Santa Barbara, CA 93106, USA}
\affiliation{Materials Department, University of California Santa Barbara, Santa Barbara, CA 93106, USA}

\author{J.T. Dong}
\affiliation{Materials Department, University of California Santa Barbara, Santa Barbara, CA 93106, USA}

\author{A.-H. Chen}
\affiliation{Université Grenoble Alpes, CNRS, Grenoble INP, Institut Néel, 38000 Grenoble, France.}

\author{M. Hocevar}
\affiliation{Université Grenoble Alpes, CNRS, Grenoble INP, Institut Néel, 38000 Grenoble, France.}

\author{C.J. Palmstrøm}
\affiliation{Electrical and Computer Engineering, University of California Santa Barbara, Santa Barbara, CA 93106, USA}
\affiliation{California NanoSystems Institute, University of California Santa Barbara, Santa Barbara, CA 93106, USA}
\affiliation{Materials Department, University of California Santa Barbara, Santa Barbara, CA 93106, USA}

\author{S.M. Frolov}
\email{frolovsm@pitt.edu}
\affiliation{Department of Physics and Astronomy, University of Pittsburgh, Pittsburgh, PA, 15260, USA}

\begin{abstract}
More and more materials, with a growing variety of properties, are built into electronic devices. This is motivated both by increased device performance and by the studies of materials themselves. An important type of device is a Josephson junction based on the proximity effect between a quantum material and a superconductor, useful for fundamental research as well as for quantum and other technologies. When both junction contacts are placed on the same surface, such as a two-dimensional material, the junction is called ``planar". One outstanding challenge is that not all materials are amenable to the standard planar junction fabrication. The device quality, rather than the intrinsic characteristics, may be defining the results. Here, we introduce a technique in which nanowires are placed on the surface and act as a shadow mask for the superconductor. The advantages are that the smallest dimension is determined by the nanowire diameter and does not require lithography, and that the junction is not exposed to chemicals such as etchants. We demonstrate this method with an InAs quantum well, using two superconductors - Al and Sn, and two semiconductor nanowires - InAs and InSb. The junctions exhibit  critical current levels consistent with transparent interfaces and uniform width. We show that the template nanowire can be operated as a self-aligned electrostatic gate. Beyond single junctions, we create SQUIDs with two gate-tunable junctions. We suggest that our method can be used for a large variety of quantum materials including van der Waals layers, topological insulators, Weyl semimetals and future materials for which proximity effect devices is a promising research avenue.
\end{abstract}

\maketitle

\section{Broad Context}

In the rapidly advancing field of quantum technologies, some of the breakthroughs are based on the better understood materials, such as silicon and aluminum~\cite{arute2019quantum, philips2022universal}. However, unresolved challenges with quantum control, coherence times, and scalability make it relevant to explore new materials for quantum devices~\cite{de2021materials}. In parallel, the development of quantum materials takes advantage of quantum devices as a way of exploring the materials properties and discovering exotic quantum states.

\section{Previous Work: Super-Semi Planar Junctions}

One prominent direction is the search and validation of Majorana zero modes in topological superconductors~\cite{frolov2020topological}. Among several approaches, planar junctions based on semiconductor quantum wells have been proposed as hosts for Majorana zero modes~\cite{pientka2017topological}. Experimental claims of topological superconductivity have been made in InAs and HgTe junctions~\cite{fornieri2019evidence, ren2019topological, dartiailh2021phase}. The possibility of entering the $\pi$-junction state at large magnetic fields have been experimentally explored~\cite{hart2017controlled, ke2019ballistic, dartiailh2021phase}. Edge state transport was detected using Josephson interference~\cite{hart2014induced, pribiag2015edge}. Super-semi junctions were also explored as a platform for transmon-like superconducting qubits~\cite{casparis2018superconducting}.

\section{Previous Work: Shadowing}

Nanowires and nanoflakes have been used to shadow nanowires, for making \textit{in situ} Josephson junctions, superconducting islands, and normal-superconductor junctions~\cite{gazibegovic2019bottom,khan2020highly,pendharkar2021parity}. Patterned membranes have also been used as shadow masks~\cite{carrad2020shadow, schuffelgen2019selective}. Finally, fabricated tall walls were used for shadow deposition of metals onto selective-area-grown structures ~\cite{heedt2021shadow,jung2021universal}. Vanadium oxide nanowires were used as etch templates to produce metallic wires~\cite{vsordan2001removable}.

Superconductor-semiconductor planar junctions to two-dimensional electron gases (2DEGs) were defined by selective etching of the superconducting layer~\cite{kjaergaard2016quantized, kjaergaard2017transparent, shabani2016two, lee2019transport, mayer2019superconducting, mayer2020gate, dartiailh2021missing}. Anodic oxidation method has been developed as one alternative to etching~\cite{drachmann2021anodic}. Though high quality junctions were demonstrated, a concern with this is that the weak link is directly exposed to potentially degrading chemical, thermal or mechanical processing. The requirement for selective etching of metal on semiconductor also limits material choices. 

\begin{figure*}
\includegraphics{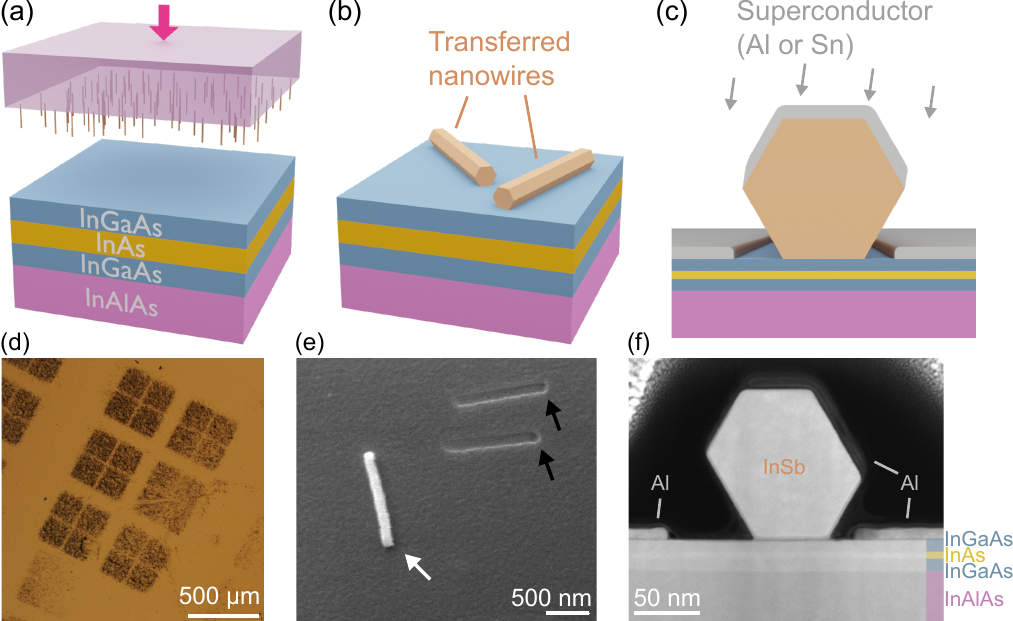}
\caption{\label{fig_illustration}
The nanowire shadow mask method.
(a) A mother chip with nanowires (top) and a 2DEG chip (bottom) are aligned to make contact with each other.
(b) After contact, nanowires adhere to the 2DEG chip due to van der Waals interaction.
(c) A thin layer of superconductor, Al or Sn, is evaporated onto the 2DEG with nanowires acting as shadow masks. Gray arrows indicate the flux of superconductor.
(d) Optical microscope image of a dummy chip after nanowire transfer. Nanowires were grown in square patterns which got imprinted to this chip. 
(e) Scanning electron microscope (SEM) image of a practice GaAs chip after Al evaporation. The white arrow indicates a nanowire. The black arrows indicate masked areas with nanowires removed by ultrasound.
(f) Cross-section high-angle annular dark-field (HAADF) image of a junction shows Al thin film interrupted by the shadowing InSb nanowire on an InAs 2DEG. The layer structure of the quantum well as well as other materials are labeled.
}
\end{figure*}

\section{Motivation and Approach}

We are motivated by extending the concept of shadowing by nanoscale objects to planar junctions. In our approach, nanowires act as shadow masks protecting part of the surface from superconductor deposition.  The key steps of creating a clean interface and the junction along the critical dimension can be performed \textit{in situ} or \textit{ex situ}, and can be extended to a variety of materials of interest. The advantages are in avoiding chemical processing of the bulk of the weak link and realizing dimensions that are challenging with liftoff.

\section{List of Results}

We deposit InAs and InSb nanowires onto shallow InAs quantum wells, and use wires as shadow masks during the deposition of Al and Sn superconductors. The nanowires are used as self-aligned gates which allows for gate-tunable supercurrents. The product of switching current and normal resistance, $I_{sw}R_N$, is 0.4 mV for Al Josephson junctions (JJs), and 0.86 mV for Sn JJs. Both of these values are higher than their corresponding superconducting gaps, suggesting high transparency interfaces. We fabricate single Josephson junctions and dc-SQUIDs using this method. While this manuscript is dedicated to the introduction of the nanowire shadow planar junction technique, in simultaneous manuscripts we report how the technique allows us to study the anomalously large second-order Josephson effects~\cite{smash_second_harmonic,zarassi_thesis} as well as missing Shapiro steps, a Majorana signature, in the trivial regime near zero magnetic field~\cite{smash_missing_shapiro}.

\section{Future Relevance}

The concept of making planar junctions using nanoscale objects can be taken beyond superconductor-semiconductor structures, and beyond nanowires. One interesting direction to explore is whether junctions can be made using van der Waals materials as weak links. Many important quantum materials are synthesized as bulk crystals rather than thin films, which complicates device fabrication. Dispersal of nanowires on the surface of a 3D crystal can take care of defining the critical dimension. Other objects, such as nanoflakes, nanonetworks or nanocubes, can be used as shadow masks.

\begin{figure*}
\includegraphics{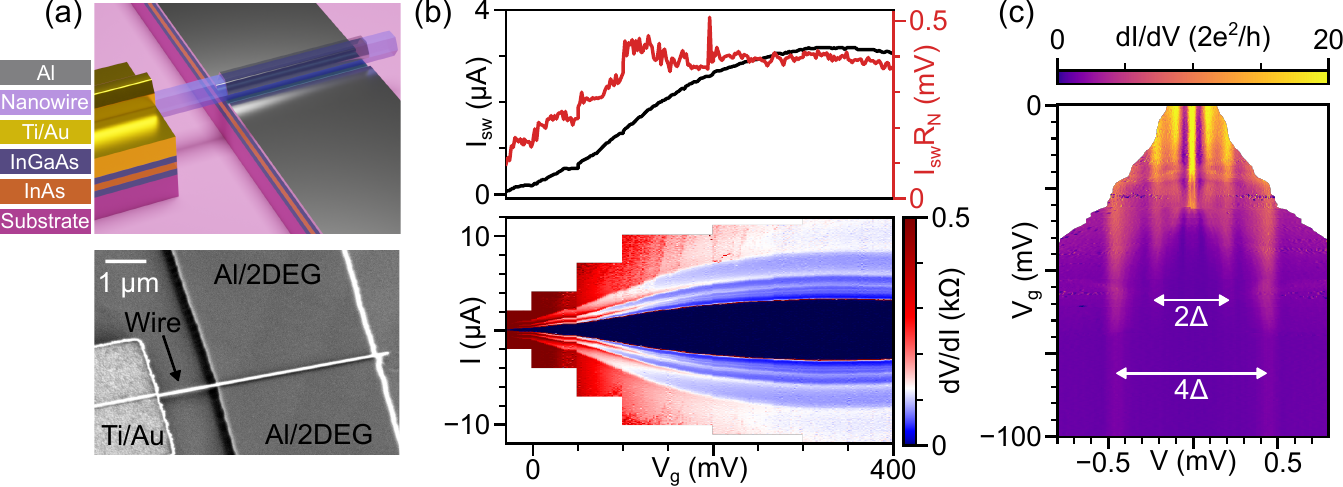}
\caption{\label{fig_Al_JJ}
Al-InAs-Al Josephson junction device (JJ-1).
(a) Schematic diagram (top) and SEM image (bottom) of a device similar to JJ-1. The nanowire is contacted by the Ti/Au electrode to function as a self-aligned gate.
(b) Differential resistance ($dV/dI$, bottom) as a function of the bias current ($I$) and the gate voltage ($V_g$). Top:  switching current ($I_{sw}$, black) and the $I_{sw} R_{N}$ product (red) extracted from data in the bottom.
(c) Differential conductance ($dI/dV$) as a function of the bias voltage ($V$) and $V_g$. White arrows indicate the first and second multiple Andreev reflections using $\Delta = 0.21$ meV.
}
\end{figure*}

\section{Figure~\ref{fig_illustration}: The Method}

The nanowire shadow-mask technique is illustrated in Fig.~\ref{fig_illustration}.
We start with a shallow InAs quantum well, with a thickness of 5 nm located 10 nm below the surface and grown on an InP substrate~\cite{lee2019transport,lee2019contribution}. The mobility of the two-dimensional electron gas (2DEG) is approximately 20,000 cm$^2$/(Vs) as
measured on a separate calibration sample. To transfer nanowires onto the 2DEG chip surface, a mother chip with as-grown vertical nanowires is positioned against the 2DEG chip (Fig.~\ref{fig_illustration}(a)). In Fig.~\ref{fig_illustration} these are InSb nanowires. The two chips are gently touched together, and nanowires stick to the 2DEG chip due to van der Waals interaction. This is shown schematically in Fig.~\ref{fig_illustration}(b). A thin layer of superconductor, which is either Al or Sn, is deposited onto the 2DEG coated with nanowires (Fig.~\ref{fig_illustration}(c)).

Panel (d) shows an optical image of a semiconductor surface with nanowires transferred. The darker squares are an imprint of the pattern in which the nanowires were grown on the mother chip. The density of transferred wires can be controlled by adjusting the growth density, or simply by imprinting several times so that the remaining density becomes smaller and smaller.

For our first generation of devices, which use InSb nanowires, the contact between the two chips is made inside an MBE vacuum cluster. This allows, in principle, to not break vacuum between the 2DEG growth, nanowire transfer and superconductor deposition. In our case, the quantum well was grown earlier and stored outside. To remove the native oxide, atomic hydrogen cleaning is performed prior to superconductor deposition~\cite{webb2015electrical,pendharkar2021parity}. For our second generation of devices, which use HfO$_x$-covered InAs nanowires, the contact is made in air to increase the efficiency without sacrificing the cleanness of the super-semi interface. 

The scanning electron microscope (SEM) image of a practice GaAs chip with nanowires coated with 10 nm of Al is shown in Fig.~\ref{fig_illustration}(e). Here, nanowires are removed by ultrasound after superconductor deposition revealing the shadow. However, for planar junctions we keep the wires in place, for two reasons. First, we use nanowires as markers of where the shadowed regions are, since the superconductor film is thin and difficult to resolve under microscope. Second, we use nanowires as self-aligned gates, see Fig.~\ref{fig_Al_JJ}.

The cross-section scanning transmission electron microscope (STEM) image of a planar junction is shown in Fig.~\ref{fig_illustration}(f). This image is reproduced in the schematic panel (c).  Al layer is capped \textit{in situ} with evaporated aluminum oxide or oxidized in the load-lock chamber to prevent dewetting. The InSb nanowire is not in electrical contact with the 2DEG due to the native oxide on the nanowire's surface: the planar junction structure consists of two Al contacts on the left and right and the quantum well. The Al film is smooth and the contacts are sharply defined. The gap of 140 nm between the left and right Al layers is determined by the nanowire width. The top and the right facets of the hexagonal nanowire are also coated with Al but that layer is not in contact with the 2DEG. The thicknesses of InGaAs/InAs/InGaAs layers are 10/5/10 nm. A 10 nm layer of Al is deposited onto the quantum well substrate cooled to 80 K.
\begin{figure*}
\includegraphics{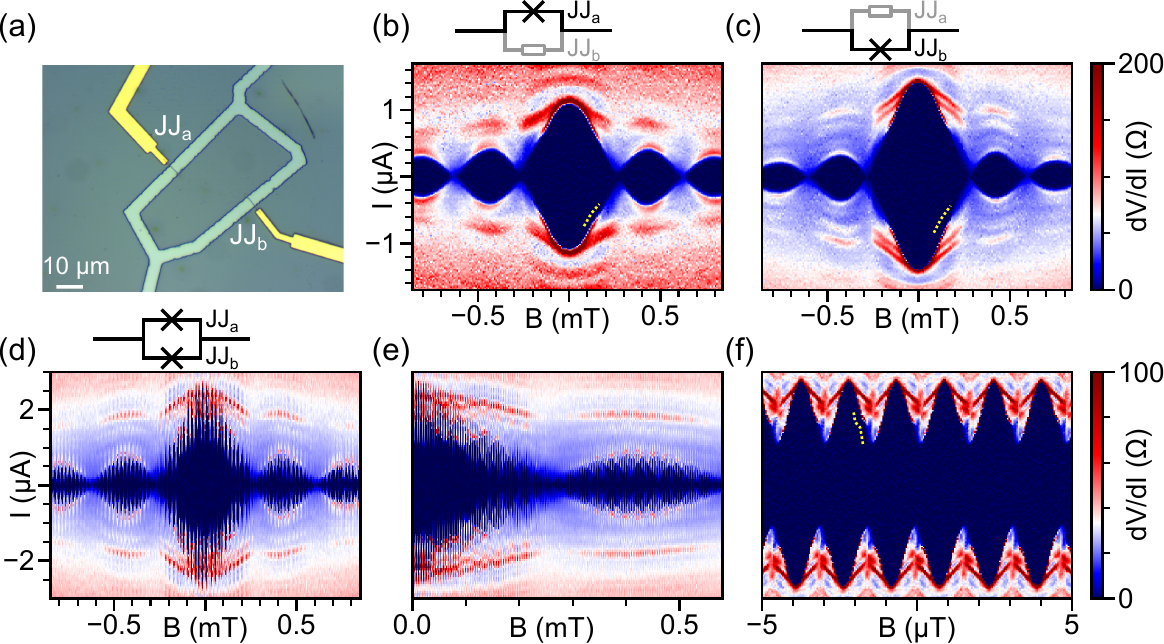}
\caption{\label{fig_Al_SQUID}
Tunable Al-2DEG SQUID (SQUID-1).
(a) Optical microscope image of SQUID-1 with junctions JJ$_a$ and JJ$_b$ labeled. A JJ is a structure shown in Fig.~\ref{fig_Al_JJ}(a).
(b) $dV/dI$ as function of the current ($I$) and the magnetic field ($B$) when JJ$_{a}$ is ``on" ($V_{g,a} = 500$ mV) and JJ$_{b}$ is ``off" ($V_{g,b} = -500$ mV).
(c) JJ$_{b}$ is ``on'' ($V_{g,b} = 500$ mV) while JJ$_{a}$ is ``off" ($V_{g,a} = -700$ mV). 
(d) Both junctions are ``on" ($V_{g,a} = V_{g,b} = 500$ mV). 
(e) and (f) narrower flux range data showing resolved SQUID oscillations.
Yellow dotted lines in panels (b), (c), and (f) highlight kinks in $I_{sw}$ modulation studied in another paper \cite{smash_second_harmonic}. Fields in (b-f) are shifted by about -0.25 mT to make low frequency Fraunhofer oscillation symmetric about $B = 0$
}
\end{figure*}

\section{Figure~\ref{fig_Al_JJ}: Al/2DEG Josephson Junction}

The final step in the junction fabrication is an etch step. As can be seen in Fig.~\ref{fig_illustration}(e) the superconductor fully surrounds the nanowire shadow. But a junction requires two separate superconducting contacts. As Fig.~\ref{fig_Al_JJ}(a) illustrates, we etch away the areas colored pink, while the superconductor and the quantum well is left in the gray/metallic area.

We leave the nanowire in place to be used as a self-aligned electrostatic gate. In order to do this, we choose InAs nanowires which are coated with hafnium oxide while still standing (prior to step shown in Fig.~\ref{fig_illustration}(a)). We use InAs in order to make sure that the nanowire itself is conducting as required for the gate effect. The reason for a hafnium oxide coat is to protect the nanowire during the wet etching step. It also acts as a dielectric layer between the nanowire gate and the quantum well. A Ti/Au contact is made to the nanowire in order to actuate the gate. Hafnium oxide is removed in the contact area by argon-ion milling. The nanowire is suspended in between the Ti/Au contact and the Al/2DEG junction. Note that the Al layer on top of the nanowire is removed during the wet etch step in the suspended segment. Transport data in the main text are taken from devices that use HfO$_x$-covered InAs nanowires. Devices using InSb wires, without hafnium oxide and without the gate contact, are presented in supplementary materials~\cite{sm}.

We present a gate-tunable Josephson supercurrent trace in Fig.~\ref{fig_Al_JJ}(b). Measurements are performed in a dilution refrigerator at temperatures of 50 mK. Supercurrent can be fully suppressed by the gate voltage near zero. Though non-superconducting current cannot be pinched off, likely due to parallel conduction in the quantum well stack. The data are nearly symmetric about $I = 0$, indicating that the junction is non-hysteretic, thus in the over-damped regime~\cite{mccumber1968effect, stewart1968current}. Josephson junctions are often characterized by the parameter $I_{sw} R_{N}$, where $I_{sw}$ is the switching current to the finite voltage state, and $R_{N}$ is the normal state resistance. Note that in a given measurement setup $I_{sw}$ may be lower than the intrinsic Josephson critical current. Extracted $I_{sw}$ and $I_{sw} R_N$ are depicted in the top panel of Fig.~\ref{fig_Al_JJ}(b).  $I_{sw}$ and $I_{sw} R_{N}$ first increase, then saturate as the gate voltage $V_{g}$ increases. $I_{sw} R_N$ saturates near $0.4$ mV, which is larger than the typical superconducting gap of Al (0.2 meV).
Larger $I_{sw} R_N$ products typically indicate higher interface transparency and more generally, an absence of major factors reducing critical currents. 

The differential resistance $dV/dI$ exibits dips alongside the center dark-blue area which are multiple Andreev reflections (MARs) (Fig.~\ref{fig_Al_JJ}(b), bottom). They become peaks sticking to constant voltages in differential conductance $dI/dV$ (Fig.~\ref{fig_Al_JJ}(c)). The $i$-th MAR contributes to peaks at $\pm 2\Delta/i \text{e}$, where $\Delta$ is the induced superconducting gap, and e is the elementary charge. We estimate $\Delta$ to be $0.21$ meV from positions of first and second resonances. This value is consistent with previous works~\cite{zhang2021evidence, mayer2019superconducting}.

\section{Figure~\ref{fig_Al_SQUID}: Al/2DEG SQUIDs}

Next, we fabricate gate-tunable superconducting quantum interference devices (SQUIDs). Fig.~\ref{fig_Al_SQUID}(a) shows SQUID-1 consisting two Al-InAs-Al planar junctions, JJ$_a$ and JJ$_b$, connected in parallel. The two nanowires have landed nearby during the transfer step. For this figure we define the ``off" state where $I_{sw}$ is zero, and the ``on" state where $I_{sw}$ is above 1 $\mu$A. However, the critical current is continuously tunable with gates similar to Fig.~\ref{fig_Al_JJ}. We note that junctions are still conducting in the ``off'' state even though supercurrent vanishes. When the magnetic field is applied in the out-of-plane direction, Fraunhofer-like modulation of $I_{sw}$ in the two junctions show similar periodicity (Fig.~\ref{fig_Al_SQUID}(b),\ref{fig_Al_SQUID}(c)), as expected for junctions with the same nominal geometry. The effective length ($L_{eff}$, along the direction of the current) of the junction can be calculated with $L_{eff} = \Phi_0 / W B_0$, where $\Phi_0 = h/2\text{e}$ is the magnetic flux quantum, $W = 5$ $\mu$m is the junction width, $B_0 = 0.28$ mT is the modulation period. We get $L_{eff} = 1.5$ $\mu$m. This value is an order of magnitude larger than the typical geometric length ($L$) of these junctions. This difference likely indicates large London penetration depth ($\lambda_{L}$). With $L_{eff} = L + 2 \lambda_{L}$, we estimate $\lambda_{L}$ close to 700 nm. The large difference between $L_{eff}$ and $L$ is also reported in other systems~\cite{rosen2021fractional}. This junction length significantly exceeding the geometric length should be taken into account when estimating system size for planar junction topological superconductivity attempts, it may be too large to enter the single mode regime~\cite{pientka2017topological, ren2019topological, fornieri2019evidence}.

When both junctions are in the ``on" state, an extra high-frequency modulation of the switching current due to SQUID interference arises [Fig.~\ref{fig_Al_SQUID}(d)-\ref{fig_Al_SQUID}(f)]. SQUID-1 area $S = \Phi_0/B_0'=1.33 \times 10^3$ $\mu$m$^2$, where $B_0' = 1.55$ $\mu$T is SQUID modulation period. This area is consistent with the designed enclosed area which is $1.28\times 10^3$ $\mu$m$^2$.

\begin{figure*}
\includegraphics{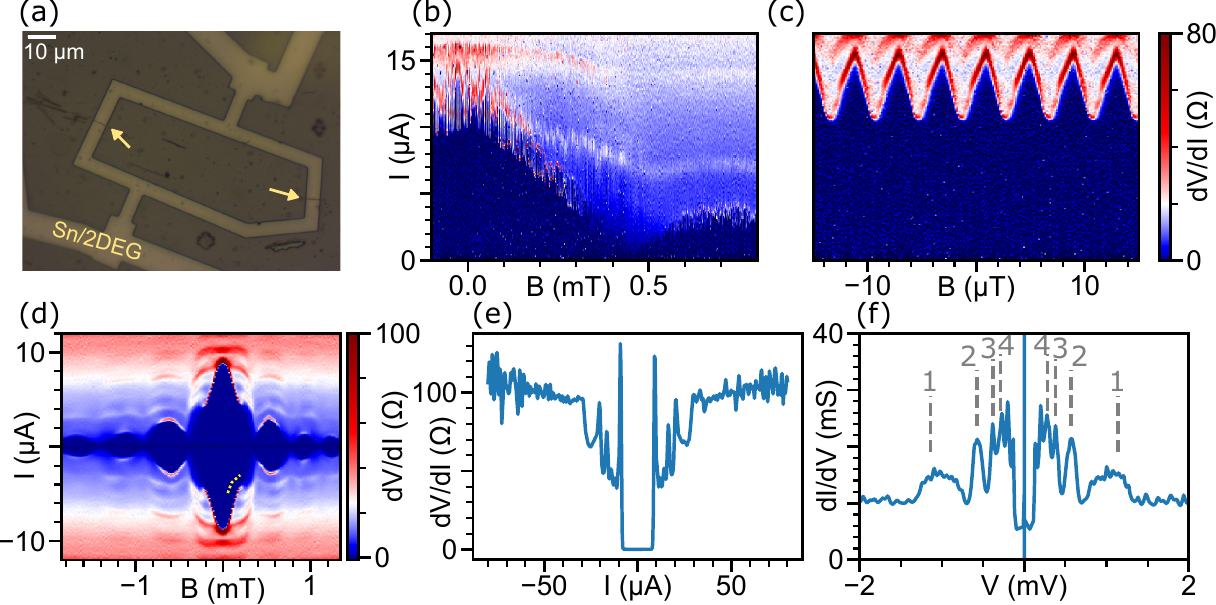}
\caption{\label{fig_Sn}
Sn/2DEG SQUID (SQUID-2) and Josephson junction (JJ-2) at 1.1 K.
(a) Optical microscope image of SQUID-2. Two shadow nanowires are indicated by arrows. Junctions are underneath nanowires.
(b) $dV/dI$ as function of the current ($I$) and the magnetic field ($B$) in SQUID-2.
(c) SQUID modulation in a narrower magnetic field range.
(d) $dV/dI$ as a function of $I$ and $B$ in JJ-2. The yellow dotted line highlights a kink in $I_{sw}$ studied in \cite{smash_second_harmonic}.
(e) Zero-field $dV/dI$-$I$ relation in JJ-2. 
(f) Zero-field $dI/dV$ as function of the voltage ($V$) in JJ-2. Dashed lines are positions of multiple Andreev reflections with indices labeled for $\Delta = 0.57$ meV. A series resistance of 1.4 $\Omega$ and a field offset of -0.26 mT are used for recalibrating data in panel (d-f)
}
\end{figure*}

\section{Figure~\ref{fig_Sn}: Sn/2DEG Devices}

The nanowire shadow-mask technique comes with a degree of flexibility in materials combinations. For instance, one constraint of wet etching is that not every metal can be selectively etched on a semiconductor. Nanowire shadowing removes the need for wet etching in the weak link. 

To demonstrate this flexibility, we use another superconductor, Sn (Fig.~\ref{fig_Sn}).  Sn has a higher critical temperature than Al (3.7 K in Sn vs 1.2 K in Al), thus a larger superconducting gap, and stronger spin-orbital coupling compared to Al. Sn has been used to demonstrate hard gap, parity preserving transport, and field resilience in nanowire superconductor-semiconductor devices~\cite{pendharkar2021parity}. These properties make Sn a promising material for realizing topological superconductors in hybrid devices. At the same time, despite our efforts, selective etching of Sn on InAs or InSb has been a challenge.

Sn is deposited onto InAs 2DEGs covered with InAs nanowires in the same setup that is used for Al deposition. The substrate is cooled to 80K for Sn deposition. Sn is capped with a layer of evaporated aluminum oxide. Two Sn/2DEG devices, SQUID-2 and JJ-2 are presented in Fig.~\ref{fig_Sn}. Measurements are made at a temperature of 1.1 K, which is much higher than the measurement temperature for Al devices. The measurement at 50 mK gives similar $I_{sw}$ and $I_{sw} R_{N}$~\cite{sm}. Sn devices are based on InAs nanowires, but the wires were not used as gates, for simplicity. Compared to Al/InAs, a different wet etch recipe is used to remove Sn and InAs outside the junction (see supplementary materials~\cite{sm}).

SQUID-2 shows similar behavior to the Al/2DEG SQUID-1 while the switching current is several times larger [Figs.~\ref{fig_Sn}(a)-\ref{fig_Sn}(c)]. The Sn/2DEG single junction device, JJ-2, also behaves similarly to its Al counterpart with a larger switching current [Fig.~\ref{fig_Sn}(d)].

The $I_{sw} R_N$ product can be deduced from $dV/dI$ versus $I$ in Fig.~\ref{fig_Sn}(e). With $I_{sw} = 9$ $\mu$A, $R_N = 95$ $\Omega$, we get an $I_{sw} R_N$ product of 0.86 mV. This is larger than the typical gap of Sn which is 0.6 meV~\cite{pendharkar2021parity}. $I_{sw} R_N$ exceeding the gap is consistent with our findings from Al/2DEG nanowire shadow junctions. 

The induced gap of Sn can be deduced from MAR peaks in the differential conductance [$dI/dV$, Fig.~\ref{fig_Sn}(f)]. We estimate $\Delta=0.57$ meV. Vertical dashed lines in Fig.~\ref{fig_Sn}(f) show calculated positions of MAR peaks with indices labeled. Similar gap sizes are observed at 50 mK and in two other devices (Fig.~\ref{figS_Sn_chip2_IV}). Note that, like in Al devices, not all peaks fit the the MAR series at higher indices ($\geq 5$).

The Fraunhofer-like pattern shows kinks near half period [Figs.~\ref{fig_Al_SQUID}(b), \ref{fig_Al_SQUID}(c), Fig.~\ref{fig_Sn}(d), more and stronger examples in supplementary materials~\cite{sm}]. The Al SQUID oscillation also shows kinks. These observations can be explained by a strong second-harmonic component in the current-phase relation. A detailed study is presented in a separate manuscript~\cite{smash_second_harmonic,zarassi_thesis}.  The inversion-symmetric character of magnetic flux diffraction patterns, more clearley seen in supplementary materials~\cite{sm}, can be explained by self-field effects.

\section{Conclusions}

We introduce a method for fabricating planar Josephson junctions using nanowires as both shadow masks and self-aligned electrostatic gates. We fabricate and characterize single junctions and SQUIDs based on Al and Sn as superconducting materials. The weak link is a 2DEG in the shallow InAs quantum well. $I_{sw}R_N$s of about 0.4 mV and 0.86 mV are observed in Al and Sn junctions, respectively, both larger than their superconducting gaps.

Our studies are split in three manuscripts, this technique paper being the first. In the second, we observe deviations from a sinusoidal current-phase relation in oscillations of $I_{sw}$ in single junctions and SQUIDs. The deviations are related to a large second-order Josephson harmonic~\cite{smash_second_harmonic}. In the third manuscript, the Shapiro step measurement is performed, showing that a series of odd Shapiro steps are missing which is related to several different origins~\cite{smash_missing_shapiro}.

\section{Data Availability}
Curated library of data extending beyond what is presented in the paper, as well as simulation and data processing code are available at~\cite{zenodo}.

\section{Duration and Volume of Study}

This study proceeded over three periods.

The first period was between September 2016 to January 2017. We proved the concept of the nanowire shadowing with dummy chips.

The second period was between August 2018 to June 2019, including sample preparation, device fabrication and measurements. We explored the first-generation devices (Al, InSb nanowire, without gates). More than 7 devices on 1 chip are measured during 2 cooldowns in a dilution refrigerator, producing about 8900 datasets.

The third period was between March 2021 to February 2022. We explored the second-generation devices, (Al or Sn, InAs/HfO$_{x}$ nanowire, with self-aligned nanowire gates). 62 devices on 6 chips are measured during 8 cooldowns in dilution refrigerators, producing about 5700 datasets.

\section{Acknowledgements}

We thank E. Bakkers, S. Gazibegovic and G. Badawy for providing InSb nanowires. We acknowledge the use of shared facilities of the NSF Materials Research Science and Engineering Center (MRSEC) at the University of California Santa Barbara (DMR 1720256) and the Nanotech UCSB Nanofabrication Facility.

\section{Funding}

Work supported by the ANR-NSF PIRE:HYBRID OISE-1743717, NSF Quantum Foundry funded via the Q-AMASE-i program under award DMR-1906325, the Transatlantic Research Partnership and IRP-CNRS HYNATOQ, U.S. ONR and ARO.

\section{References}

\bibliographystyle{apsrev4-1}
\bibliography{references.bib}

\clearpage
\beginsupplement

\centerline{\textbf{\large{Supplementary Materials: Planar Josephson Junctions Defined by Nanowire Shadowing}}}

\section{Author Contributions}

A.-H.C and M.H. grew InAs nanowires.
M.P., J.S.L., C.D., A.M., S.D.H., B.S., J.T.D., and C.J.P grew quantum wells and superconducting films with nanowire shadowing.
S.T. performed STEM study.
P.Z., A.Z., B.Z., S.M., and H.W. fabricated devices.
P.Z., L.J, V.V.d.S, and A.Z. performed measurements.
P.Z. and S.M.F. wrote the manuscript with inputs from all authors. S.M.F. directed the project.



\section{Device information}

\begin{table}[H]
\caption{\label{tab:info_dev} Information of devices discussed in this manuscript. JJ-1, JJ-2, SQUID-1, and SQUID-2 are discussed in the main text. The rest are discussed in supplementary materials}
\begin{ruledtabular}
\begin{tabular}{llllll}
Chip name & JJ-\# & SQUID-\# & Chip reference & Wafer & Description\\
\hline
Al-chip-1 & S3-S8 & - & 2019 2DEG & SH180919 &  SH180919 QW + InSb NWs\\
Al-chip-2 & S9-S11 & - & 20210329 Al InAs 2DEG & MP738 & MP728 QW + NW637 InAs NWs\\
Al-chip-3 & 1, S1-S2 & 1 & 20210924 Al InAs 2DEG & MP738 & MP728 QW + NW637 InAs NWs\\
Sn-chip-1 & S12-S16 & S1 & 20210907 Sn InAs 2DEG & MP739 & MP728 QW+ NW637 InAs NWs\\
Sn-chip-2 & 2, S17-S20 & 2, S2 & 20211009 Sn InAs 2DEG & MP739 & MP728 QW + NW637 InAs NWs\\
\end{tabular}
\end{ruledtabular}
\end{table}

\section{Full Methods}

\begin{figure}[H]\centering
\includegraphics[width=\linewidth]{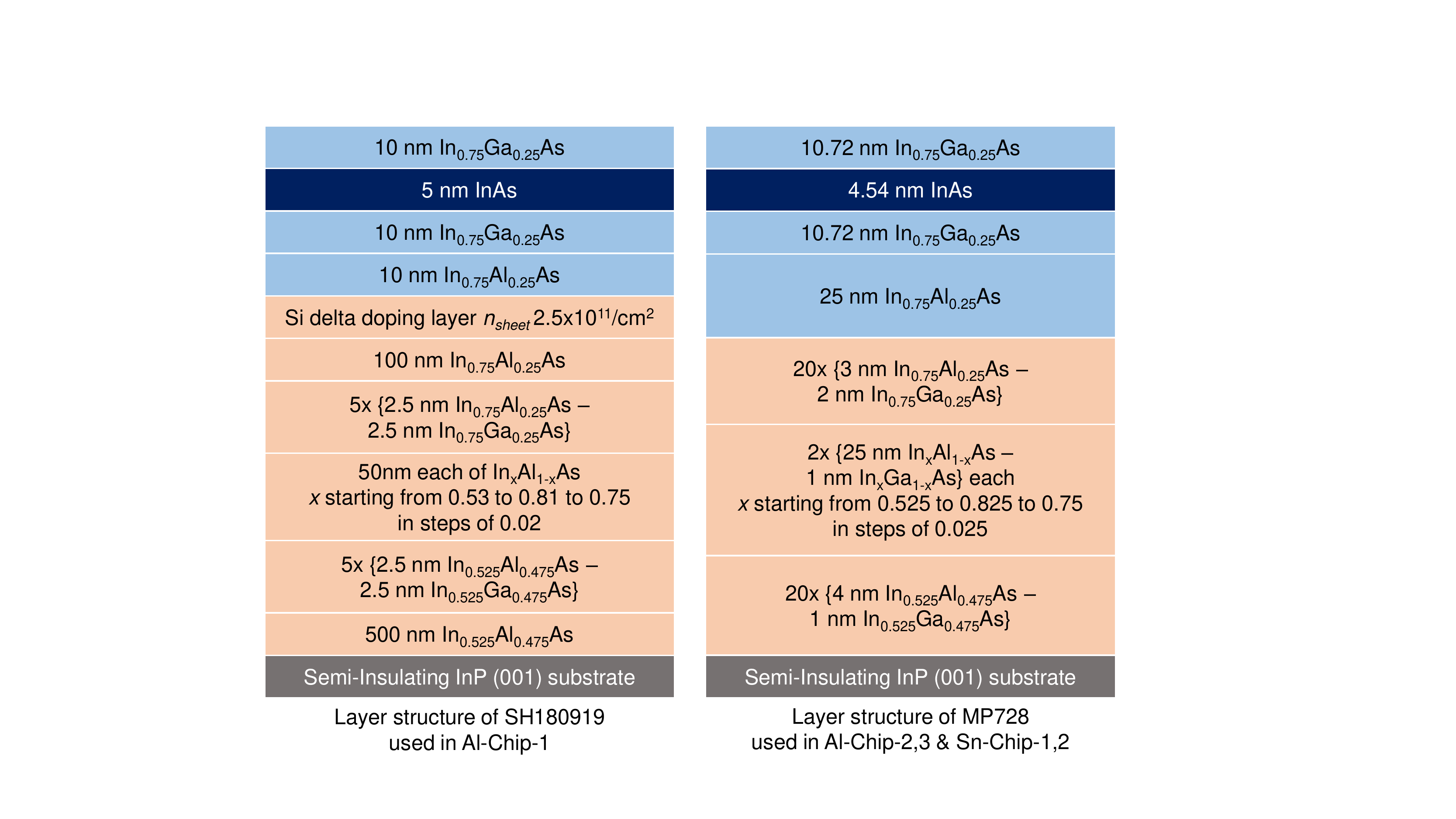}
\caption{\label{figS_qw_structures}
Layer structures of quantum well samples used in this manuscript.
}
\end{figure}

\textbf{2DEG growth}.
The InAs quantum wells are grown by molecular beam epitaxy (MBE) similar to work reported in~\cite{lee2019transport,lee2019contribution}. The layer structures of the quantum well samples are depicted in Fig.~\ref{figS_qw_structures}. For SH180919, the mobility on a calibration sample, without the nanowire and superconductor, was estimated to be approximately 25,000 cm$^2$/Vs at a sheet carrier density of $1 \times 10^{12}$~cm$^{-2}$. This corresponds to an electron mean free path of approximately 400~nm. For a similar calibration of MP728, the mobility was measured to be 20,000 cm$^2$/Vs at a sheet carrier density of $4.6 \times 10^{11}$~cm$^{-2}$, corresponding to an electron mean free path of approximately 200~nm. Electron mobility and density in the underlying 2DEG are expected to change after nanowires are brought in contact with the 2DEG surface and superconductors are evaporated.


\textbf{Nanowire growth (InAs)}. 
The InAs nanowires were grown by MBE using the gold assisted VLS mechanism. The samples were prepared as follow. InAs substrates (111)B and (001) were wet deoxidized using NH$_4$OH during 5~min then rinsed in DI water. Gold colloids were immediately deposited after deoxidation on the susbtrates and samples were introduced to the MBE chamber. After an annealing step at 500~$^\circ$C under As vapour, nanowires were grown at 420~$^\circ$C by opening the In shutter. After 75 min of growth, the In shutter was closed and the sample cooled down to room temperature under vacuum. InAs wires under those conditions (BEP V/III ratio 20 and In flux $5\times 10^{-7}$ Torr) exhibit 60 nm diameter, have a reduced tapered and measure above 4 um in length. Once removed from the MBE chamber, the samples were introduced to an ALD system for HfO$_2$ deposition (120 $^\circ$C, 150 cycles, about 5 nm).

\textbf{Superconducting layer}. 
The InAs quantum wells were removed from UHV post growth of the quantum well and were put face down, i.e. “smashed”, on the surface of their corresponding InAs nanowire chips (for devices with InSb nanowires, the smash was performed in an MBE machine in prior to the hydrogen cleaning step described below). This smashing of the two surfaces, leads to part of the nanowires being transferred to the quantum well chip due to electrostatic attraction. It was observed that smashing the two chips such that the quantum well chip was facing down lead to less debris from the smash being left behind on the same chip. Since this same quantum well chip, now with the nanowires, would be used for superconductor evaporation and subsequent device fabrication, maintaining optimum cleanliness would be beneficial.

While this process of smashing the chips could in principle be carried out without breaking UHV, execution of the same remains a part of future work. Advantages of conducting this process in UHV include preventing the oxidation of the surface of the quantum well while also enabling the growth of the nanowires in UHV (in a separate run) or pre-cleaning of the nanowire chip with atomic hydrogen upon insertion in UHV, prior to smashing.

The nanowire+quantum well “smashed” chip was then re-introduced in UHV for subsequent surface preparation, evaporation of superconductor and passivation. Atomic hydrogen cleaning was performed at a substrate temperature of about 475-480~$^\circ$C, as measured by a thermocouple, under constant rotation, at a chamber pressure of $5 \times 10^{-6}$~Torr of primarily hydrogen. The samples once cooled to room temperature were moved to a neighboring UHV chamber equipped with a cryogenic sample stage operating at about 80~K (cooled with liquid nitrogen) and effusion cells of aluminum and tin. The samples were then pre-cooled for over an hour in direct contact with the cryogenic sample stage at 80~K, before evaporation of aluminum or tin was started. Typical nominal growth rates of about 10~nm/hr were used for evaporation of the superconducting films of 10~nm thickness at oblique incidence. Al-Chip-1 was immediately moved out of vacuum to oxidize the Al film. Al-Chip-2,3, and Sn-Chip-1,2, were immediately coated with a protective layer of electron-beam evaporated AlOx (nominally 3~nm), before the samples could warm up to room temperature. Further details of this process are described in Ref.~\cite{pendharkar2021parity}.

\textbf{Device fabrication (Al/2DEG)}.
First, the chip is pre-patterned with coordinate markers by e-beam lithography (EBL), e-beam evaporation of Ti/Au, and lift-off.
Second, optical-microscope images of nanowires are taken for device designing.
Third, gate electrodes are made by EBL, Al etching (CD26 developer:water 1:20 for 2 min, or Transene aluminum etchant type D at 50 $^\circ$C for 10 s), Ar ion milling to remove HfO$_x$ (500 V, 25 mA for 60 s in a Plassys system), and then 10/120 nm Ti/Au by e-beam evaporation.
At last, mesas of Josephson junctions and SQUIDs are made by EBL using negative tone resistor ma-N 2403, followed by Al etching and InAs etching (water:1M citric acid:38\% H$_3$PO$_4$:H$_2$O$_2$ 220:55:3:3 for 8 min, with agitation). Etch with the Al etchant again to avoid undercuts after 2DEG etching.
The ratios of solutions are in volume. EBL resists are ``baked" in a vacuum chamber for 8 hours or longer at room temperature to avoid heating.

\textbf{Device fabrication (Sn/2DEG)}.
The fabrication of Sn/2DEG devices is similar to that of Al/2DEG. In some devices, a 10/100 nm Ti/Au layer is deposited over the leads to short possible discontinuities in the Sn film and avoid cracks due to wire bonding. This Ti/Au layer is made by EBL, Ar ion milling of Al$_2$O$_3$ (250 V, 15 mA for 90 s in a Plassys system), and e-beam evaporation. The Sn layer is etched by first developing in ma-N 525 for 3 min to remove unexposed ma-N 2403 and Al$_2$O$_3$, then etching in 36.5\%-38\% HCl:water 1:1000 for 10 s. We find the etching rate of Sn in the diluted HCl solution difficult to control. Large undercuts of several $\mu$m under the resist may happen. We carefully check the result under an optical microscope after etching for every 5 s to avoid possible undercuts.

\section{Uncertainty in the $I_{sw}$ $R_{N}$ product}

$I_{sw} R_N$ is widely used for characterizing a Josephson junction. We find it is sensitive to the method we use for extracting parameters. For example, in the top-middle part (near 150 mV, 10 $\mu$A) of Fig.~\ref{fig_Al_JJ}(b) bottom panel, there are red tilted lines which is due to a background increasing linearly against the current. This background may be a result of the gating effect from the source-drain voltage. The larger the current bias, the larger the $R_N$ extracted. This explains the discontinuity in the $I_{sw} R_N$ curve at $V_g = 100$ and $200$ mV because here $R_N$ is extracted at the largest current bias in each vertical line. In the Sn device, we also observe that $R_N$ increases by about $15\%$ between 35 $\mu$A and 80 $\mu$A, in both positive and negative sides [Fig.~\ref{fig_Sn}(e)]. This may be due to the heating effect at large currents. We take $R_N=95$ $\Omega$ which is the value near the switching current around $\pm 30$ $\mu$A to minimize the uncertainty caused by heating.

\section{Series resistance and multiple switching currents}

\begin{figure}[H]\centering
\includegraphics{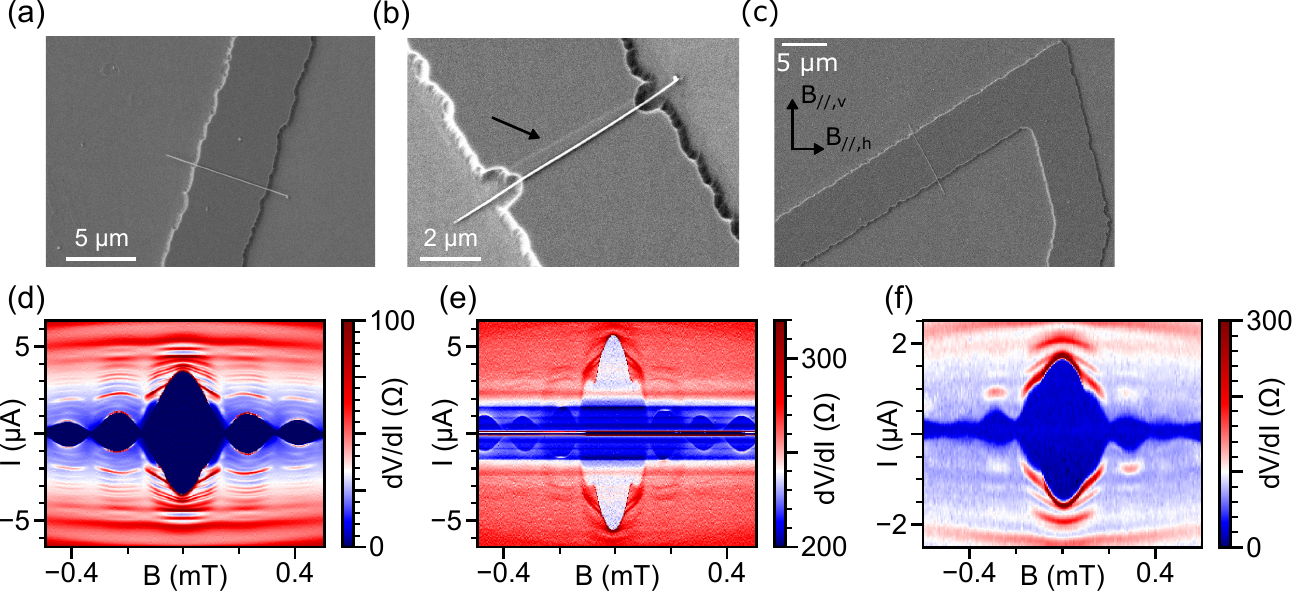}
\caption{\label{figS_second_batch_JJ}
Data from multiple devices on Al-chip-2.
(a) SEM of a typical device. The mask nanowires are InAs coated with HfO$_x$ protecting layer. These nanowires are resistant to the 2DEG etchant. No gate electrodes are made on this chip.
(b) SEM of a device showing a rare event that the nanowire is accidentally moved during the fabrication. The arrow shows the junction which is a bright thin line across the mesa.
(c) SEM of device JJ-S9. Directions of vertical and horizontal in-plane magnetic fields are labeled.
(d-f) Fraunhofer diffraction patterns in JJ-S9, JJ-S10, JJ-S11, respectively. JJ-S10 manifests extra superconducting switchings which may be due to breaks of Al film in the leads. Offsets ($\leq 0.2$ mT) are applied to the field so that $I_{sw}$ maximizes at zero field.
}
\end{figure}

\begin{figure}[H]\centering
\includegraphics{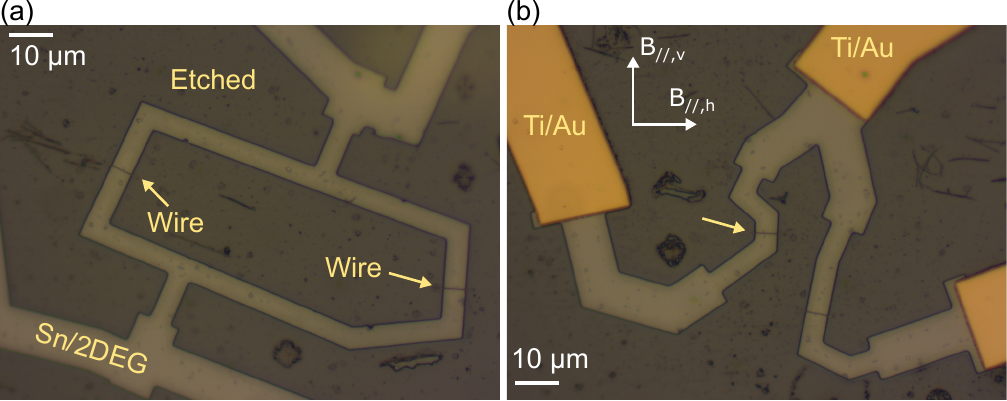}
\caption{\label{figS_Sn_images}
Optical microscope image of Sn/2DEG devices in the main text. Both devices are on Sn-chip-2. (a) SQUID-2. This panel duplicates Fig.~\ref{fig_Sn}(a). (b) JJ-2 which is indicated by the yellow arrow.
}
\end{figure}

Devices are mostly designed to be 2-terminal with long leads on chips. We bond 4 wires to each 2-terminal device to perform 4-terminal measurements, with every two wires sharing one lead. This may lead to a series resistance and multiple superconducting switching currents if a lead has non-superconducting regime besides the junction. Breaks in the superconducting film may be caused by unexpected masking nanowires on leads, scratches on the surface, or damage made by wire bonding. We have observed series resistance and multiple switching currents in both Al and Sn devices. In Al devices these issues are very rare (Fig.~\ref{figS_second_batch_JJ}). The Al film seems to be less fragile against scratches and wire bonding comparing to the Sn film. Another property of Sn that may contribute to these issues is that Sn has two common phases, one of which is not superconducting. To suppress the series resistance and extra superconducting switchings in Sn devices, we either make four-terminal leads on the chip or short long leads by a Ti/Au layer (Fig.~\ref{figS_Sn_images}). 

\section{Magnetic field offset}

Numerical offsets are applied to the magnetic field so that $I_{sw}$ always maximizes at ``zero" magnetic field. The offset is typically of the order of hundreds $\mu$T. It may be due to fluxes trapped in devices or the magnet itself.

\begin{figure}[H]\centering
\includegraphics{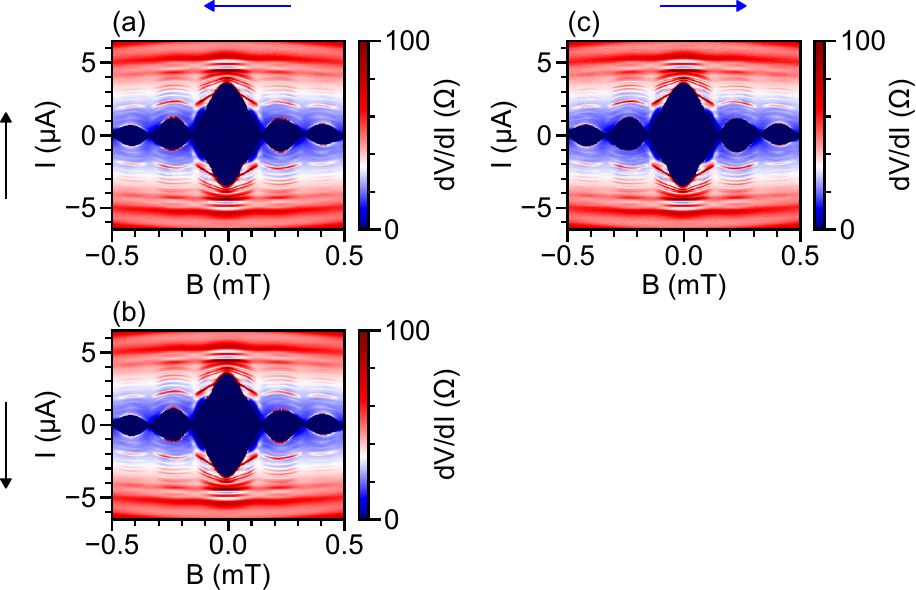}
\caption{\label{figS_Al_hysteresis}
The differential resistance $dV/dI$ in JJ-S9 shows no obvious hysteresis against the field or the current. The scan direction of the field and the current is indicated by arrows on the top and left, respectively. Original magnetic field without an offset is shown.
}
\end{figure}

\begin{figure}[H]\centering
\includegraphics{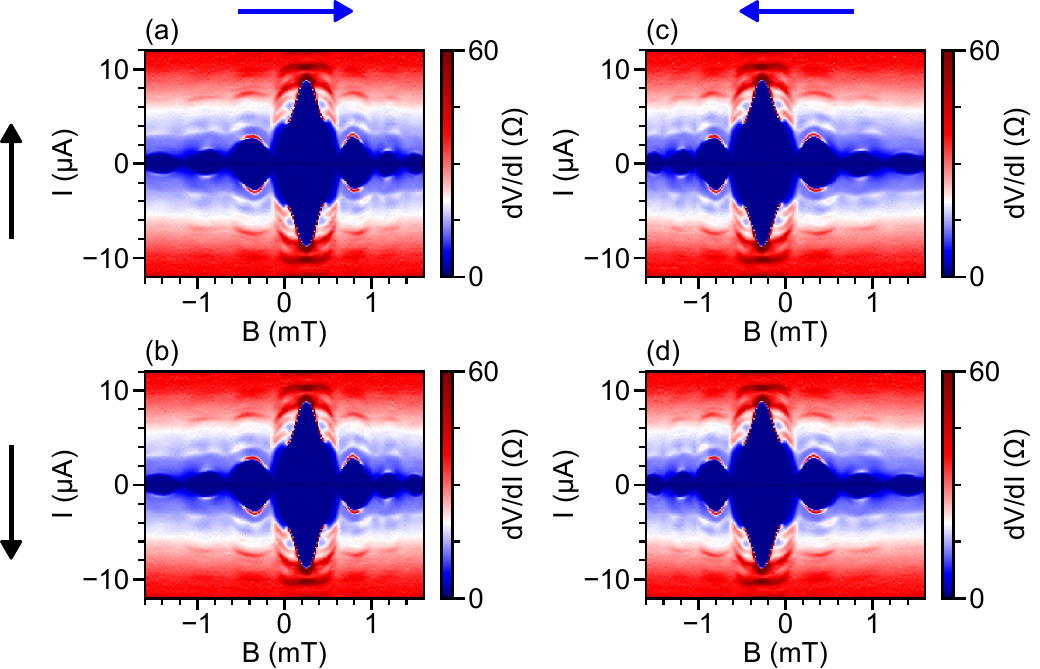}
\caption{\label{figS_Sn_hysteresis}
The differential resistance dV/dI in JJ-2 shows no obvious hysteresis about the current but obvious hysteresis about the field.  The scan direction of the field and the current bias is indicated by arrows on the top and left, respectively.
}
\end{figure}

Comment on data in Figure~\ref{fig_Sn} - Sn/InAs junctions. Devices in this batch are hysteretic about the magnetic field. The field is scanned from negative to positive in Fig.~\ref{fig_Sn}(d). The left part of the oscillation is stretched while the right part of the oscillation is squeezed (For details, see Fig.~\ref{figS_Sn_hysteresis}). The hysteretic behavior may be caused by fluxes trapped in junctions. Non-linear behavior of a magnet at mT range may also lead to a hysteresis. However, we do not observe such a strong hysteretic behavior about the magnetic field in Al devices in the same fridge (Fig.~\ref{figS_Al_hysteresis}). Optimization to the device design such as avoiding sharp kinks in a path can be made in future works to avoid possible vortex accumulation. 

\section{Supplementary data from Al/2DEG devices at 50 mK}

\begin{figure}[H]\centering
\includegraphics{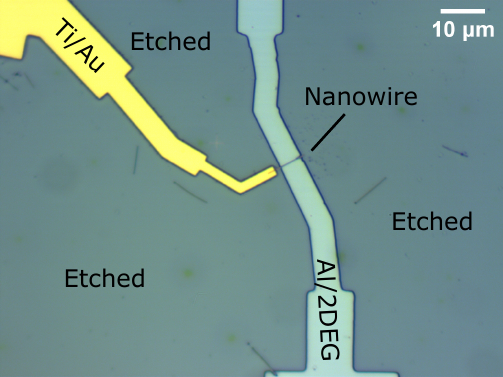}
\caption{\label{figS_JJ1_image}
Optical microscope image of JJ-1 (Al/2DEG) which is discussed in the main text. The vertical bright gray path is the Al/2DEG mesa. The yellow lead is the Ti/Au electrode contacting the nanowire which is a dark line on the top of the mesa. The rest is wet etched.
}
\end{figure}

\begin{figure}[H]\centering
\includegraphics{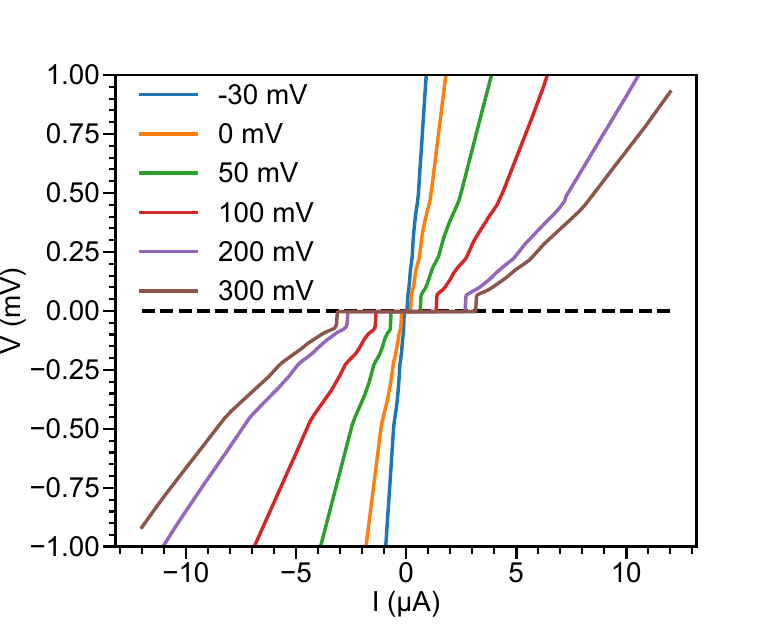}
\caption{\label{figS_JJ1_IV}
V-I curves at different gate voltages in JJ-1. Data are extracted from Fig.~\ref{fig_Al_JJ}(b).
}
\end{figure}


\begin{figure}[H]\centering
\includegraphics{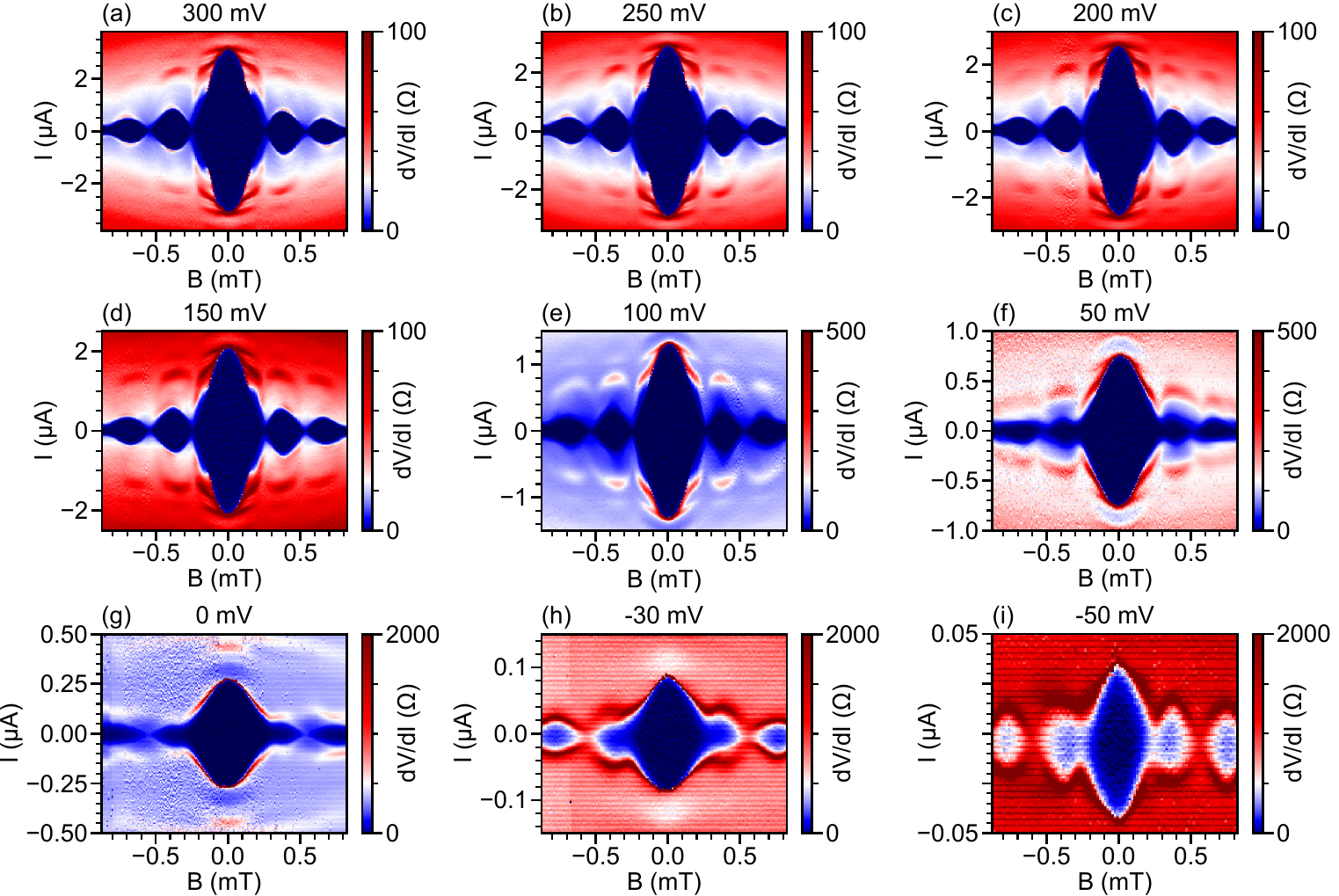}
\caption{\label{figS_JJ_1_more_Fraunhofer}
Fraunhofer diffraction patterns of JJ-1 (Al/2DEG) at a variety of gate voltages which are noted at the top of each panel. The field is shifted by -0.23 mT in all panels so that $I_{sw}$ maximizes at zero field.
}
\end{figure}

\begin{figure}[H]\centering
\includegraphics{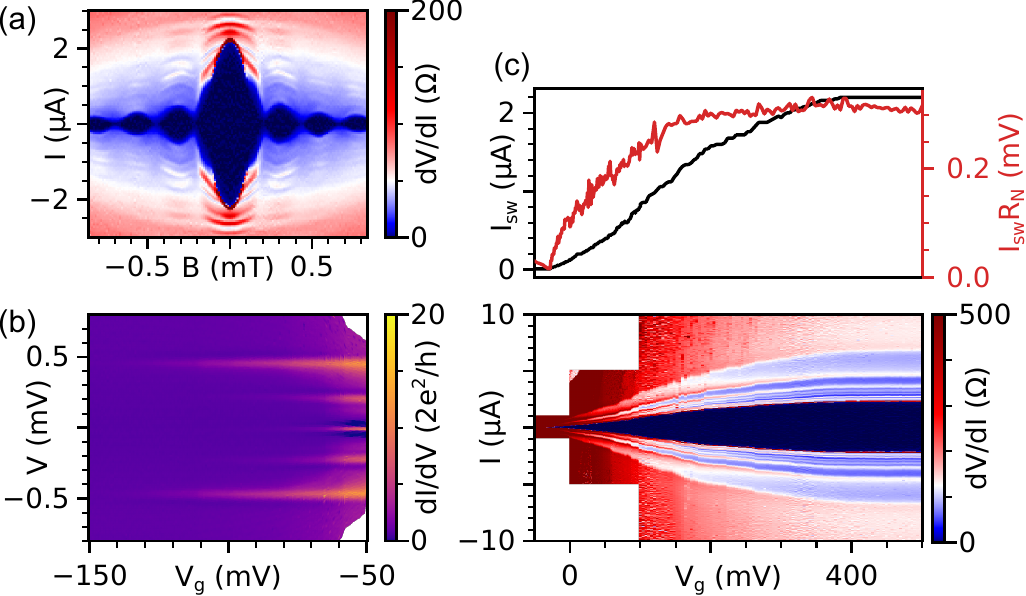}
\caption{\label{figS_JJ_S1}
Characterization of device JJ-S1 (Al/2DEG) which is similar to JJ-1 in the main text.
(a) Differential resistance as a function of the current and the magnetic field. The superconducting switching current $I_{sw}$ undergoes a Fraunhofer-like oscillation. Kinks appear in the oscillation at the half period. The gate voltage $V_{g} = 400$ mV.
(b) Differential conductance as a function of the bias voltage and the gate voltage. Horizontal resonances are due to multiple Andreev reflections. 
(c) Differential resistance as a function of the current and the gate voltage (bottom). Extracted $I_{sw}$ and $I_{sw} R_N$ as a function of the gate voltage (top). $I_{sw} R_N$ saturated near 0.3 mV.
The field is offset by a value of about -0.2 mT so that the zero field is where the switching current maximizes.
}
\end{figure}

\begin{figure}[H]\centering
\includegraphics{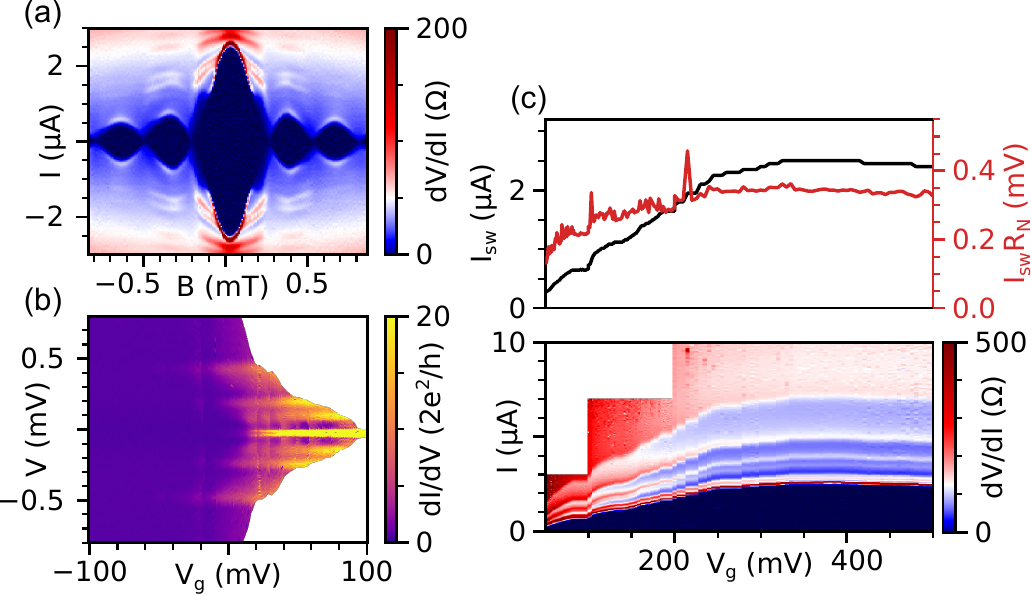}
\caption{\label{figS_JJ_S2}
Characterization of device JJ-S2 (Al/2DEG) which is similar to JJ-1 in the main text.
(a) Differential resistance as a function of the current and the magnetic field. The superconducting switching current $I_{sw}$ undergoes a Fraunhofer-like oscillation. Kinks appear in the oscillation at the half period. The gate voltage $V_{g} = 400$ mV.
(b) Differential conductance as a function of the bias voltage and the gate voltage. Horizontal resonances are due to multiple Andreev reflections. 
(c) Differential resistance as a function of the current and the gate voltage (bottom). Extracted $I_{sw}$ and $I_{sw} R_N$ as a function of the gate voltage (top). $I_{sw} R_N$ saturated near 0.35 mV.
The field is offset by a value of about -0.2 mT so that the zero field is where the switching current maximizes.
}
\end{figure}

\begin{figure}[H]\centering
\includegraphics{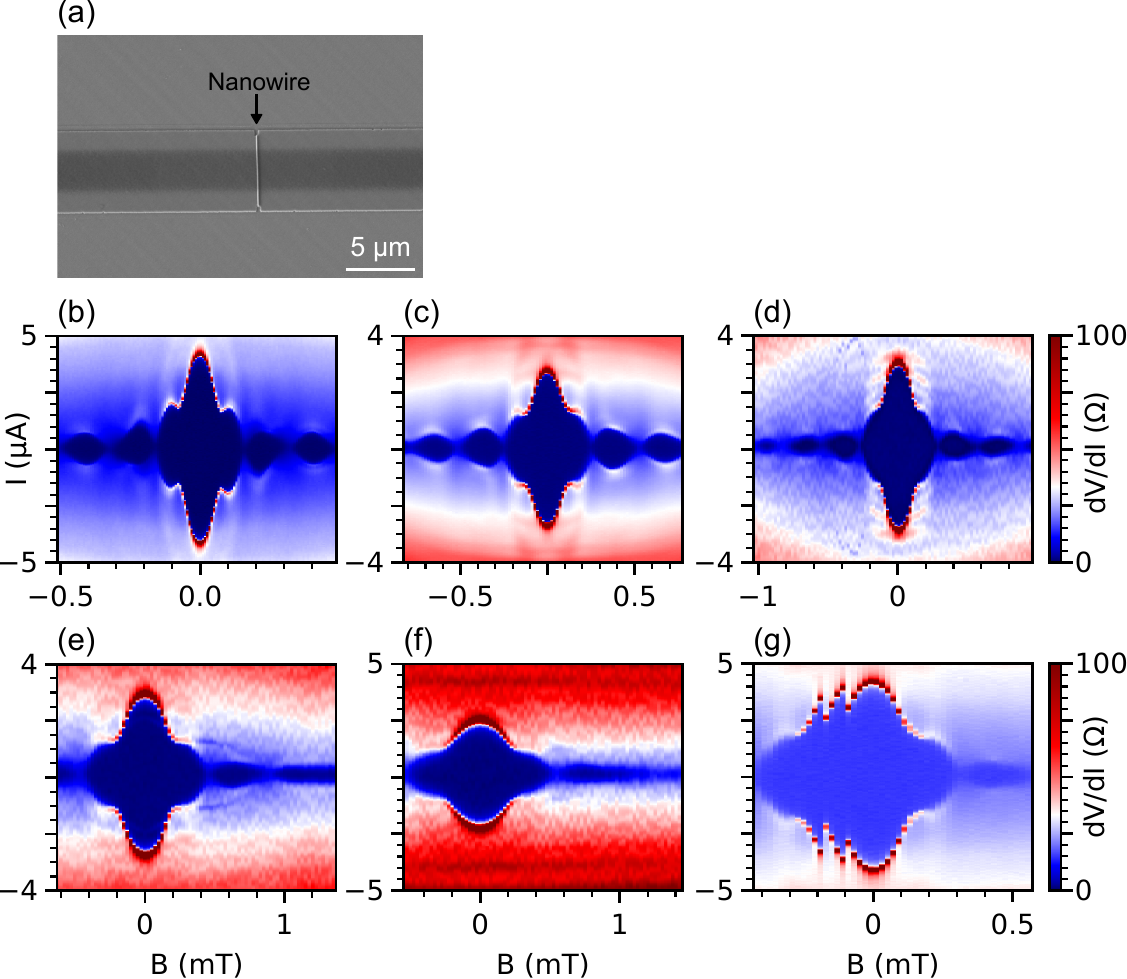}
\caption{\label{figS_first_batch_JJ}
Data from multiple devices on Al-chip-1. 
(a) SEM of a typical device on this chip. The dark stripe through the mesa is over-dosed ma-N 2403 resist residue. The designed dose in this area is accidentally doubled. The mask wire on this chip are bare InSb nanowires. They get etched by the 2DEG etchant. On newer chips we use InAs nanowires covered by HfO$_x$ protecting layer to avoid etching so that nanowires can be connect by Ti/Au electrodes next to the mesa.
(b-g) Fraunhofer diffraction patterns in JJ-S3, JJ-S4,..., JJ-S8, respectively. Offsets ($< 0.5$ mT) are applied to the field so that $I_{sw}$ maximizes at zero field.
}
\end{figure}

\begin{figure}[H]\centering
\includegraphics{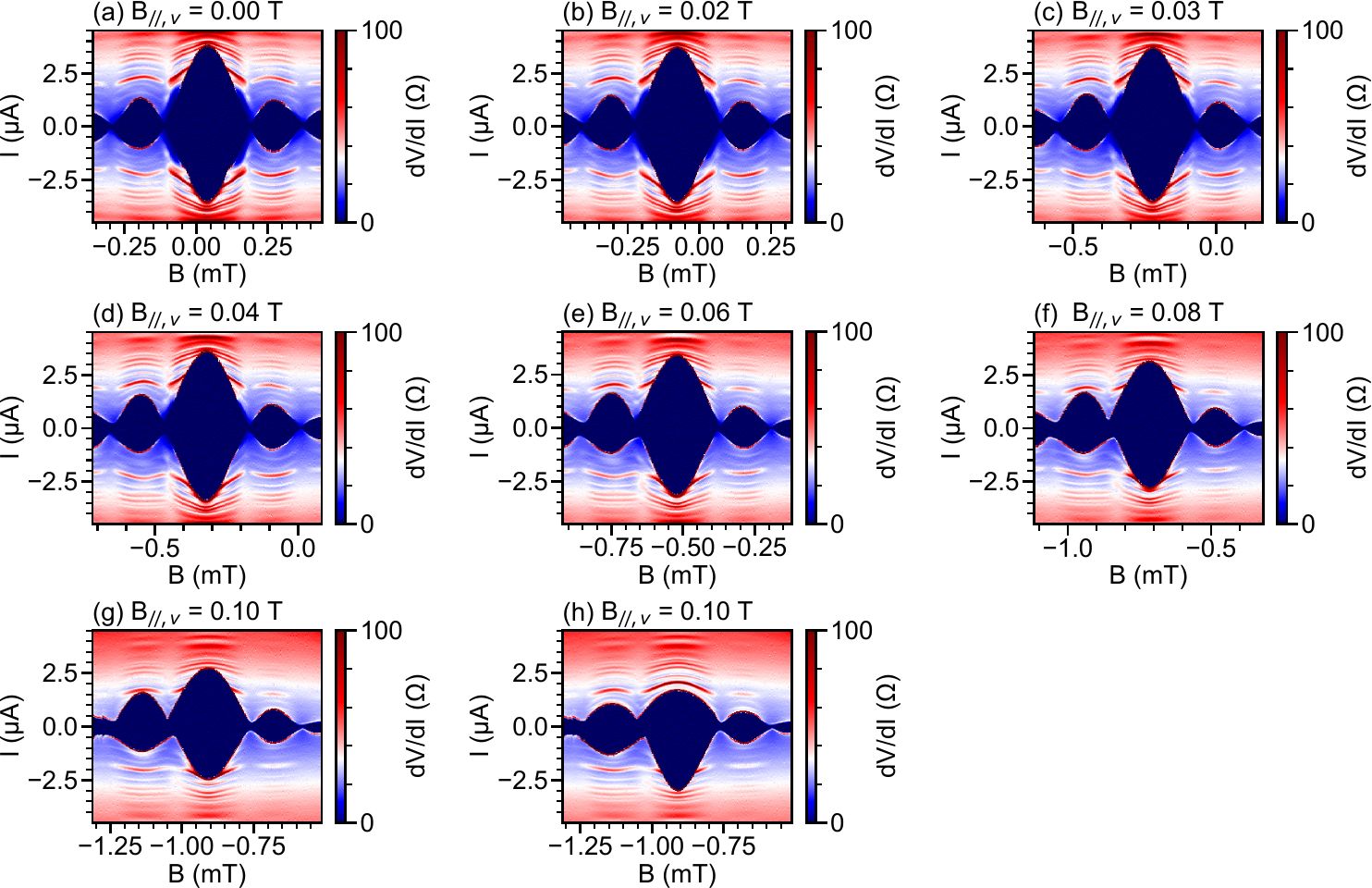}
\caption{\label{figS_Al_Bz}
(a-g) Diffraction patterns of JJ-S9 in a variety of vertical in-plane magnetic fields $B_{//,v}$ [Fig.~\ref{figS_second_batch_JJ}(c)]. The out-of-plane field where $I_{sw}$ maximizes shifts as $B_{//,v}$ increases, indicating a small misalignment between the field and device plane. The current is scanned from negative to positive.
(h) Similar as (g) but the current is scanned from positive to negative. $I_{sw}$ shows strong hysteresis in the presence of an in-plane field.
}
\end{figure}


\section{Supplementary data for Sn/2DEG devices at 50 mK and 1.1 (or 1.2) K}

\begin{figure}[H]\centering
\includegraphics{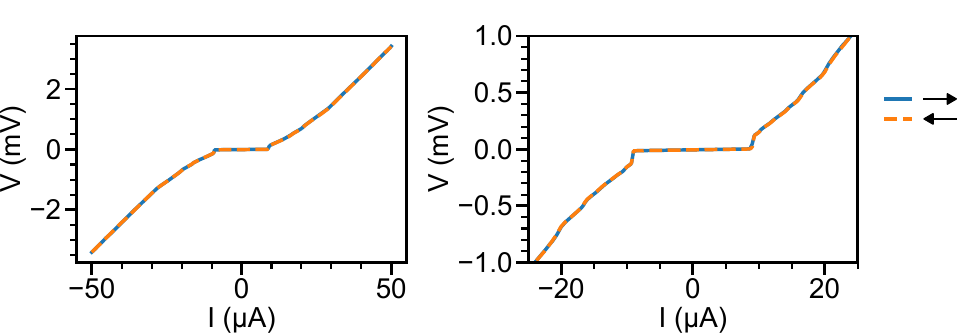}
\caption{\label{figS_JJ2_IV}
V-I curve in JJ-2. The right panel shows zoomed-in regime of the left panel. The sweep direction is indicated by black arrows on the right.
}
\end{figure}


\begin{figure}[H]\centering
\includegraphics{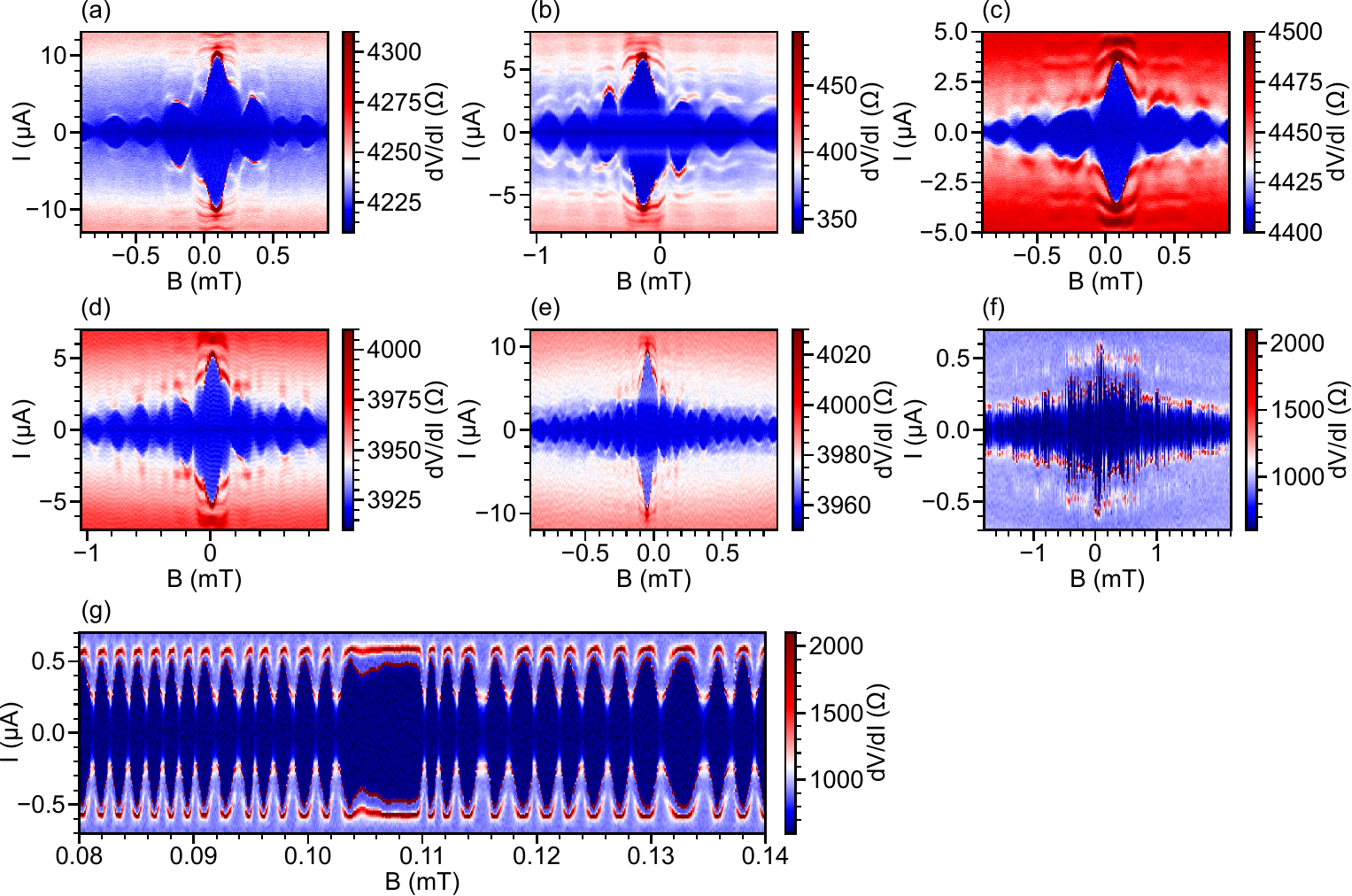}
\caption{\label{figS_Sn_chip1_1d2K}
Data from multiple devices on Sn-chip-1 which is not shown in the main text. $T = 1.2$ K.
(a-e) Single Josephson junction devices JJ-S12, JJ-S13, ..., JJ-S16. The measurement is performed with a locin-in amplifier. The AC excitation is 100 nA, the frequency is either 77.77 Hz or 88.77 Hz for different measurements. The ac signal is accidentally damped by $15\%$ due to the grounding and filters. (f-g) $dV/dI$ in SQUID-S1 measured by the DC method. JJ-S13 and SQUID-S1 are measured by the 4-terminal method. Other devices are measured by the 2-terminal method with an ac probe resistance of 3.4 k$\Omega$ (smaller than the dc resistance due to the damping). The resistance below $I_{sw}$ may be larger than the probe resistance due to non-superconducting parts in device leads and bonding pads. We suppress the on-chip series resistance by covering leads with a layer of Ti/Au or making four leads instead of two on Sn-chip-2 (Fig.~\ref{figS_Sn_images},\ref{figS_Sn_chip2_1d1K},\ref{figS_Sn_chip2_50mK}).
}
\end{figure}

\begin{figure}[H]\centering
\includegraphics{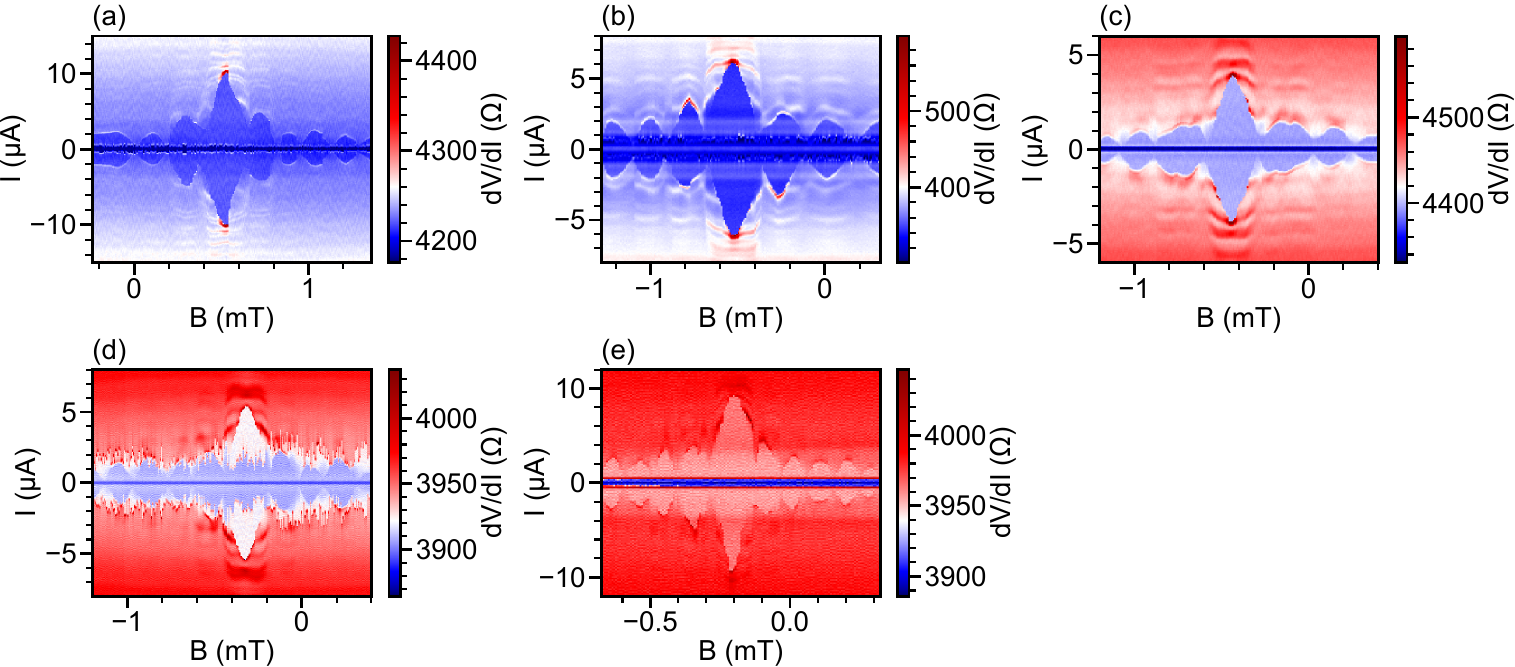}
\caption{\label{figS_Sn_chip1_50mK}
(a-e) Similar to Fig.~\ref{figS_Sn_chip1_1d2K}(a-e) except that the temperature $T = 50$ mK. Extra superconducting switches are clear at this temperature. They may be due to non-superconducting parts in leads or pads of devices. We suppress extra superconducting switches by covering leads with a layer of Ti/Au or making four leads instead of two on Sn-chip-2 (Fig.~\ref{figS_Sn_images},\ref{figS_Sn_chip2_1d1K},\ref{figS_Sn_chip2_50mK}).
}
\end{figure}

\begin{figure}[H]\centering
\includegraphics{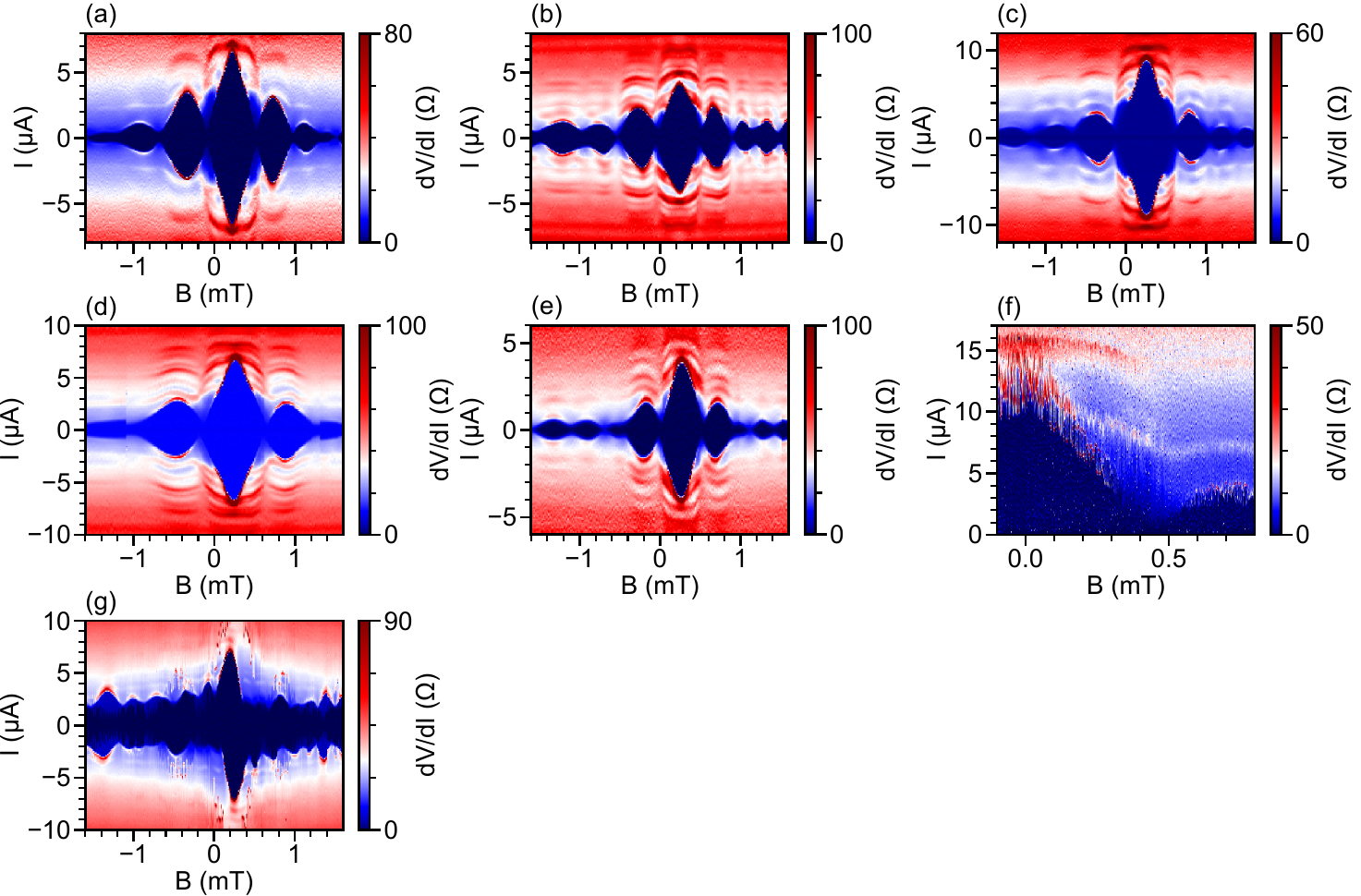}
\caption{\label{figS_Sn_chip2_1d1K}
Data from multiple devices on Sn-chip-2. $T = 1.1$ K.
(a) JJ-S17, (b) JJ-S18, (c) JJ-2, (d) JJ-S19, (e) JJ-S20, (f) SQUID-2, (g)SQUID-S2. All devices are measured with dc signal and the 4-terminal method.
}
\end{figure}

\begin{figure}[H]\centering
\includegraphics{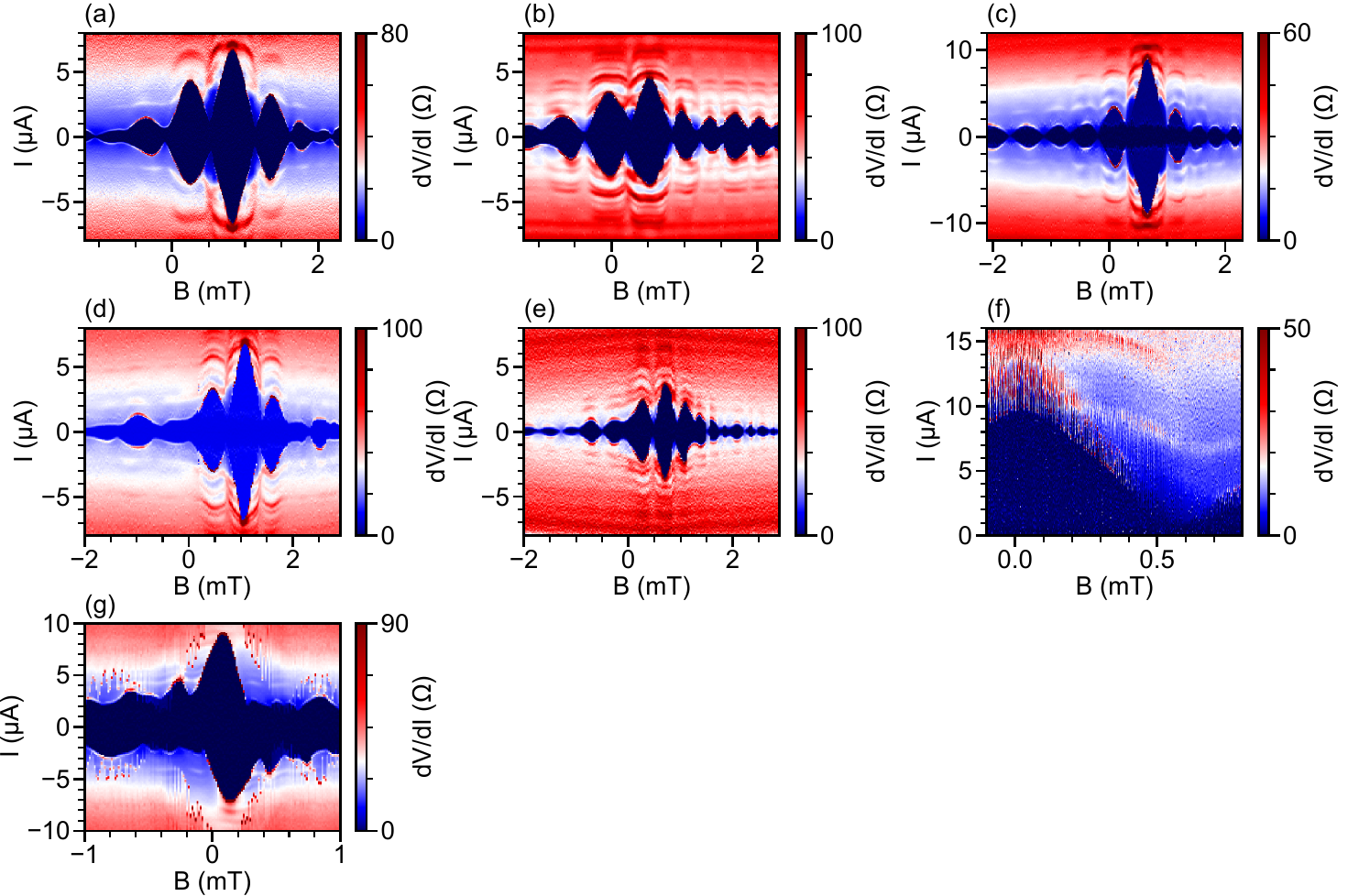}
\caption{\label{figS_Sn_chip2_50mK}
Similar to Fig.~\ref{figS_Sn_chip2_1d1K}, at 50 mK.
}
\end{figure}


\begin{figure}[H]\centering
\includegraphics{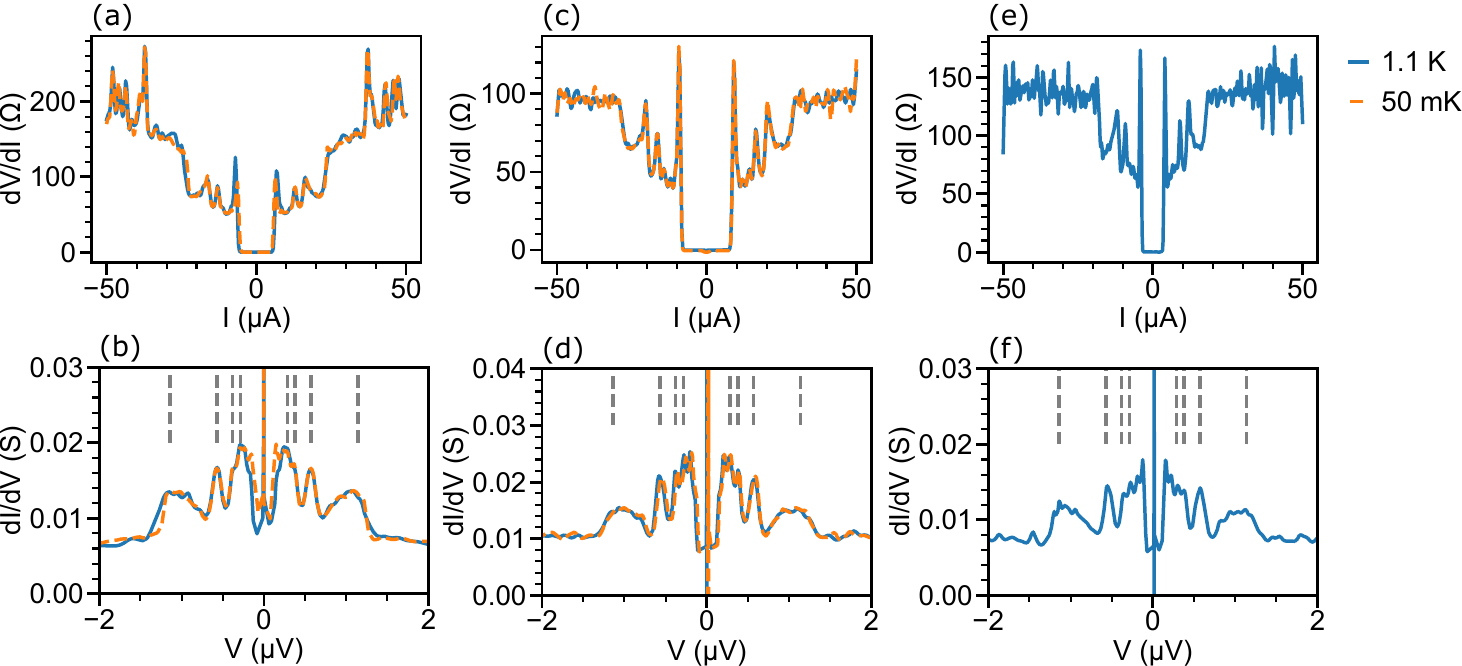}
\caption{\label{figS_Sn_chip2_IV}
Differential resistance $dV/dI$ versus current $I$ (top panels) and differential conductance $dI/dV$ versus voltage $V$ (bottom panels) for devices (a,b) JJ-S17, (c,d) JJ-2, and (e,f) JJ-S20, at zero field. The switching current $I_{sw}$ and the normal resistance $R_N$ at 1.1 K are similar to those at 50 mK. The ($I_{sw}$, $I_{sw} R_N$) extracted from (a,c,e) are (6.4 $\mu$A, 0.90 mV), (9.0 $\mu$A, 0.86 mV) , and (4.0 $\mu$A, 0.54 mV), respectively. In panel (a), $R_N$ keeps increasing above 25 $\mu$A, which may be due to the heating effect. We take $R_N = 140$ $\Omega$ which is the value near the jump at 25 $\mu$A in (a). The vertical dashed lines in (b,d,f) are calculated multiple Andreev reflection voltages at $\pm 2\Delta/i\mathrm{e}$, with $\Delta = 0.57$ meV for all three panels and $i = 1, 2, 3, 4$. Fields are shifted (see Figs.~\ref{figS_Sn_chip2_1d1K},\ref{figS_Sn_chip2_50mK}) so that the zero field is where the switching current maximizes. A series resistance of 1.4 $\Omega$ is subtracted from (c, d).
}
\end{figure}

\end{document}